\documentclass[achemso,reprint]{revtex4-1}

\usepackage[T1]{fontenc}
\usepackage{color}
\usepackage{amsmath}
\usepackage{amssymb}
\usepackage{graphicx}
\usepackage{esint}
\usepackage{lineno}
\usepackage{lipsum}

\definecolor{samblue}{rgb}{0.0,0.2,0.5}
\definecolor{samgreen}{rgb}{0,0.5,0.0}
\definecolor{samred}{rgb}{0.66,0,0}

\newcommand{\vc}[1]{{\mathbf{#1}}}
\renewcommand{\vec}[1]{{\mathbf{#1}}}

\def \<{\langle}
\def \>{\rangle}

\usepackage[normalem]{ulem}
\makeatletter

\renewcommand{\fnum@figure}{\textbf{Figure~\thefigure}}
\usepackage{units}
\usepackage{lipsum}
\usepackage{babel}
\usepackage{braket}

\renewcommand{\vec}[1]{{\mathbf{#1}}}

\makeatother

\usepackage{babel}
\begin{document}

\title{Optical Signatures of Moir\'e Trapped Biexcitons}

\author{Samuel Brem}
\email{brem@uni-marburg.de}
\author{Ermin Malic}
\affiliation{Department of Physics, Philipps University, 35037 Marburg, Germany}

\begin{abstract}
Atomically thin heterostructures formed by twisted transition metal dichalcogenides can be used to create periodic moir\'e patterns. The emerging moir\'e potential can trap interlayer excitons into arrays of strongly interacting bosons, which form a unique platform to study strongly correlated many-body states. In order to create and manipulate these exotic phases of matter, a microscopic understanding of exciton-exciton interactions and their manifestation in these systems becomes indispensable. Recent density-dependent photoluminescence (PL) measurements have revealed novel spectral features indicating the formation of trapped multi-exciton states providing important information about the interaction strength. In this work, we develop a microscopic theory to model the PL spectrum of trapped multi-exciton complexes focusing on the emission from moir\'e trapped single- and biexcitons. Based on an excitonic Hamiltonian we determine the properties of trapped biexcitons as function of twist angle and use these insights to predict the luminescence spectrum of moir\'e excitons for different densities. We demonstrate how side peaks resulting from transitions to excited states and a life time analysis can be utilized as indicators for moir\'e trapped biexcitons and provide crucial information about the excitonic interaction strength. 
\end{abstract}

\maketitle
%-------------------------------------------------------------------------
%--------------------INTRODUCTION-----------------------------------------
%-------------------------------------------------------------------------
\section*{Introduction}
\begin{figure}[ht!]
\includegraphics[width=\columnwidth]{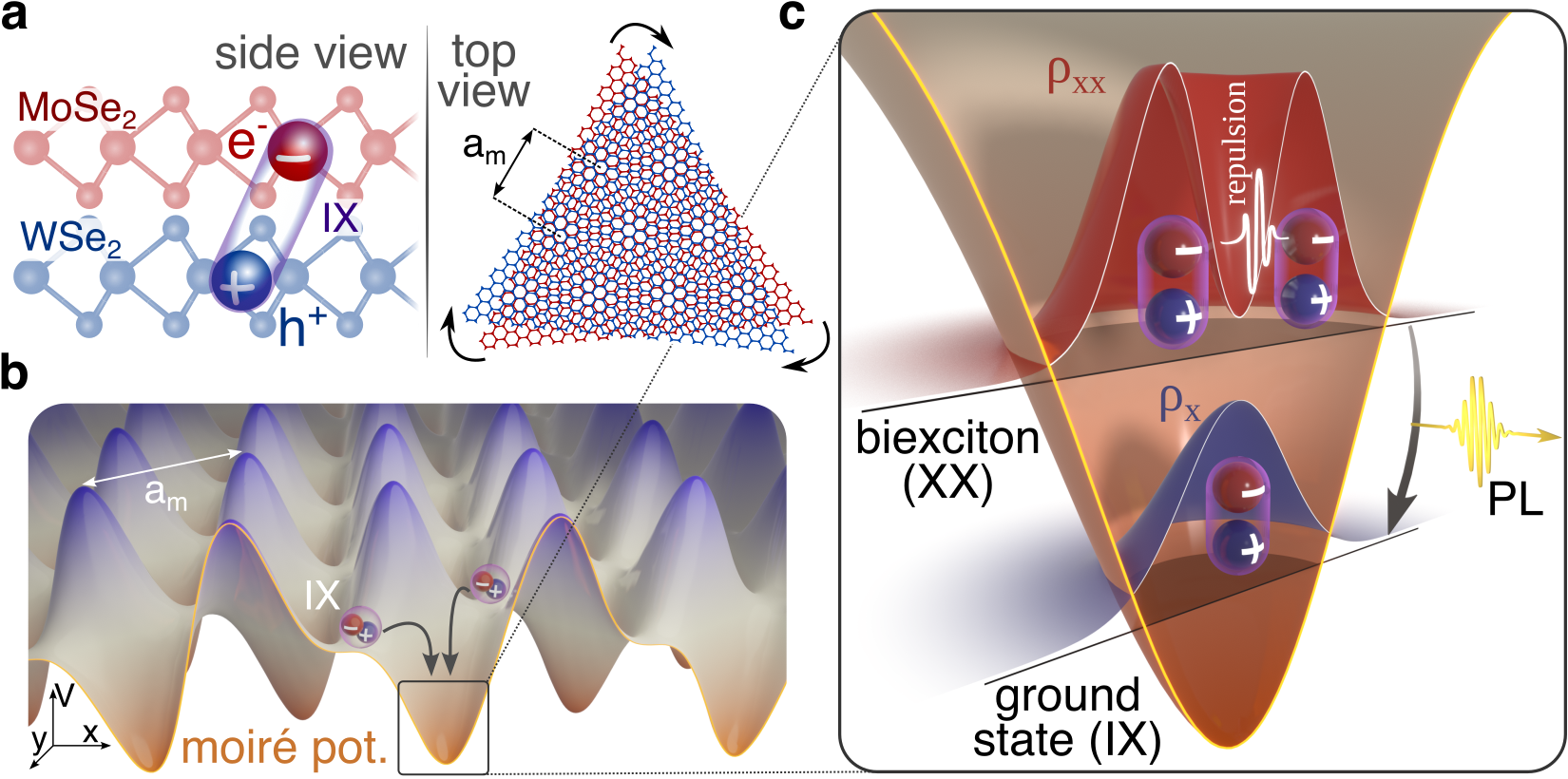}
\caption{{\bf{a}} In MoSe$_2$/WSe$_2$ bilayers the charge-separated interlayer exciton (IX) is the energetically most favourable electron-hole configuration. When the two monolayers are stacked with a small twist angle a tunable moir\'e pattern is formed with a spatially periodic potential. {\bf{b}}  The moir\'e potential captures excitons in its minima creating arrays of localized excitons. At increased densities several excitons can fill the same site forming repulsive biexcitons. {\bf{c}} The strong repulsion between  excitons leads to an efficiently altered relative motion, with important consequences for the biexciton life time and their PL spectrum.}
\label{fig:scheme} 
\end{figure}

Artificially stacked heterobilayers of transition metal dichalcogenides (TMDs) create tunable moir\'e superlattices hosting long-ranged periodic energy landscapes for electrons \cite{shabani2021deep, zhang2017interlayer, lu2019modulated}. Moreover, the optical excitation of type-II heterostructures, such as MoSe$_2$/WSe$_2$, leads to the ultrafast formation of charge-separated electron-hole pairs  \cite{jin2018ultrafast, merkl2019ultrafast, schmitt2022formation, meneghini2022ultrafast}. These long-lived interlayer excitons become trapped in potential minima of the moir\'e superlattice \cite{yu2017moire,brem2020tunable,huang2022excitons} and thereby form an array of localized bosons, whose confinement can be tuned through the stacking angle (Fig. \ref{fig:scheme}a). 

Due to their permanent out-of-plane dipole moment interlayer excitons exhibit a strong repulsive interaction \cite{nagler2017interlayer, yuan2020twist, kremser2020discrete, erkensten2022microscopic, tagarelli2023electrical}, which combined with their spatial confinement enables the formation of strongly correlated bosonic states, ranging from Mott-like to superfluid states \cite{wu2015theory,ma2021strongly, gotting2022moire,gu2022dipolar, troue2023extended}. Moreover, doping densities can be precisely controlled in 2D materials via electrostatic gating, further expanding the phase-space of feasible correlated quantum states to strongly interacting Bose-Fermi mixtures \cite{shimazaki2020strongly, zeng2023exciton}. However, an important precursor for the control of such many-body quantum states is a microscopic understanding of the interplay between moir\'e trapping and exciton-exciton interaction.

Recent photoluminescence (PL) measurements \cite{choi2020moire,wang2021diffusivity} on MoSe$_2$/WSe$_2$ have revealed intriguing density dependent changes in the transport properties of moir\'e excitons suggesting an interaction-induced delocalization at high densities \cite{brem2023bosonic}. Moreover, a series of successive PL peaks emerging above the ground state has been observed in WSe$_2$/WS$_2$ when the exciton density is increased \cite{park2023dipole}. This has been suggested to reflect the injection of several excitons into a single moir\'e site (Fig. \ref{fig:scheme}b), where the peak shifts correspond to the energy resulting from exciton repulsion. This phenomenon can be qualitatively explained within a simple Hubbard model \cite{park2023dipole}, where the multiple occupation of the same site is connected to an energy cost described by the on-site interaction $U$. However, for a more quantitative analysis of density dependent moir\'e exciton spectra, a theory beyond the Hubbard model becomes necessary. It has been shown that the strength of the dipole repulsion is comparable with the moir\'e potential \cite{gotting2022moire}, such that interaction-induced effects can lead to significant changes in the many-body wave function, which can not be captured with perturbation theory as in the Hubbard model. 

In this work, we develop a microscopic and material-specific model to compute the energies and wave functions of moir\'e excitons as well as moir\'e trapped biexcitons, i.e. two repelling excitons occupying the same moir\'e site (Fig. \ref{fig:scheme}c). Here, we set up an effective bosonic Hamiltonian for excitons parameterized via first principles calculations to calculate the eigenstates of the relative motion between two trapped excitons as a function of twist angle. We use these insights to predict amplitudes and energies of possible biexciton decay channels and show how the luminescence spectrum evolves as a function of density. In particular, we demonstrate how characteristic side peaks and a life time analysis can be utilized as an unambiguous indicator for the presence of moir\'e trapped biexcitons. Our model provides general insights into the recombination dynamics of repulsive biexcitons and can be applied to other systems of trapped bosons.    

In the following we describe the basic aspects of the theoretical model with further details provided in the Appendix. Afterwards we present the predictions of the model about biexciton properties and their experimentally accessible spectral signatures for the concrete case of an MoSe$_2$/WSe$_2$ heterobilayer.

\section*{Theoretical Model}
In this work we focus on R-stacked MoSe$2$/WSe$_2$ heterostructures, in which interlayer excitons at the K point of the Brillouin zone are the energetically most favourable electron-hole configuration \cite{lu2019modulated}. Moreover, we consider an optical excitation with circularly polarized light, such that the spin-valley degree of freedom is fixed to bright excitons in the K valley \cite{cao2012valley, mak2012control}. We set up an excitonic Hamiltonian with bosonic operators $b^{(\dagger)}_\vc{Q}$ creating (annihilating) interlayer excitons with the center-of-mass momentum $\vc{Q}$. Here we take into account the exciton dispersion and the effective moir\'e potential derived from first principle calculations \cite{brem2020tunable}. Moreover, we account for the exciton-photon coupling, which includes the material-specific selection rules of circularly polarized light for interlayer transitions at the K-point \cite{wu2018theory, yu2017moire, brem2020tunable}. Finally, we also take into account the interaction between excitons. Here, we exclude electron-hole exchange interactions due to the small overlap of electron and hole wave function for interlayer excitons \cite{erkensten2021exciton}. Moreover, the strong repulsive direct interaction between interlayer excitons will keep them at large enough distances, such that the short ranged electron-electron (hole-hole) exchange interaction can be neglected, which is further discussed in the results section. More details on the parameterization of matrix elements are given in the Appendix. 

We limit our consideration to the regime of small twist angles ($\theta_\text{twist}<1^\circ$), which allows for a simplified treatment of the many-body problem. First of all, small twist angles $<1^\circ$ correspond to moir\'e periods that are at least one order of magnitude larger then the Bohr radius of interlayer excitons ($a_\text{B}\approx 1.5$nm \cite{schmitt2022formation}). Consequently, we can neglect form factors resulting from the internal structure of the exciton and treat them as effective point particles. Secondly, for small twist angles the lowest moir\'e exciton bands are flat \cite{brem2020tunable, choi2021twist}. In this case the hopping between neighbouring moir\'e cells is strongly suppressed and the superlattice is well described as a thermodynamic ensemble of disconnected traps filled with a countable number of excitons. In this case the excitonic part of the density matrix can be expanded in multi-exciton eigenstates, 
\begin{eqnarray} \label{eq:multi-X-eigenstates}
\ket{\nu}=\sum_{\vc{Q}_1..\vc{Q}_{N_\nu}} \Psi^\nu_{\vc{Q}_1..\vc{Q}_{N_\nu}} b^\dagger_{\vc{Q}_1}..b^\dagger_{\vc{Q}_{N_\nu}}\ket{0},
\end{eqnarray}
where $\nu$ is a compound index containing the number of particles $N_\nu$ and all quantum numbers determining the configuration of the $N_\nu$-body state characterized by the Schrödinger equation $H\ket{\nu}=E_{\nu}\ket{\nu}$. In order to determine the PL signal in quasi-equilibrium, we solve the von Neumann equation within a Markov approximation, which for the emission perpendicular to the bilayer yields the PL intensity,
\begin{eqnarray} 
I_\sigma(\omega)&=&\dfrac{2\pi}{\hbar}\sum_\vc{Q}|g^{\perp}_{\sigma}(\vc{Q})|^2 J_\vc{Q}(\omega), \label{eq:PL_main} \\
J_\vc{Q}(\omega)&=&\sum_{\nu_\alpha\nu_\beta}|\bra{\nu_\alpha}b_\vc{Q}\ket{\nu_\beta}|^2P_{\nu_\beta}\delta(E_{\nu_\beta}-E_{\nu_\alpha}-\hbar\omega),\;\;\;
\end{eqnarray}
where the exciton-photon matrix element $g^{\perp}_{\sigma}(\vc{Q})$ accounts for the coupling between excitons and $\sigma$-polarized photons propagating in perpendicular direction to the bilayer. Moreover, $J_\vc{Q}(\omega)$ can be identified as the spectral intensity of excitons (Lehman representation) depending on statistical weights $P_\nu=\text{exp}[-(E_\nu-\mu_x N_\nu)/(k_BT)]/Z_x$.
In order to evaluate Eq. \ref{eq:PL_main} for different fillings, i.e. different chemical potentials, we determine the wave functions and eigenenergies of multi-exciton complexes localized in a moir\'e trap. Here we limit our consideration to the case of one and two excitons per site, since previous works suggest that the inter-excitonic repulsion leads to a delocalization at larger fillings \cite{brem2023bosonic}. When solving the one- and two-exciton Schrödinger equation, we use a harmonic approximation of the moir\'e potential around its minima $M(\vc{r})\approx m_x \omega_0^2 r^2/2-U_0$, yielding the confinement frequency
\begin{eqnarray} \label{eq:confinement_energy}
\omega_0=\dfrac{2\pi}{a_m}\sqrt{\dfrac{2U_0}{3m_x}},
\end{eqnarray}
determined by the exciton mass $m_x$, the depth of the moir\'e potential $U_0$ and its wavelength $a_m$, cf. Fig. \ref{fig:scheme}a. Within the harmonic trapping potential we can find an analytical form of the single exciton states as well as the center-of-mass motion of two excitons. In contrast, the relative motion between two excitons is additionally affected by the inter-exciton repulsion, such that the corresponding eigenvalue problem is solved numerically. Details on the calculation are given in the Appendix.

Finally, we obtain the emission signal from one and two exciton states,
\begin{eqnarray} \label{eq:PL_final}
I_\sigma(\omega)&=&\dfrac{2\pi}{\hbar}\sum_{\nu_1}\bigg(|g^\perp_{\sigma}(\nu_1)|^2 P^{(1)}_{\nu_1}\delta(E^{(1)}_{\nu_1}-\hbar\omega) \nonumber\\
&+&\sum_{\nu_2}\Gamma^\sigma_{\nu_2\nu_1} P^{(2)}_{\nu_2}\delta(E^{(2)}_{\nu_2}-E^{(1)}_{\nu_1}-\hbar\omega)\bigg).
\end{eqnarray}
Here, the superscripts $(1)$ and $(2)$ indicate single and two-particle energies/occupation probabilities, respectively, and the relevant eigenstates are referred to via the quantum numbers $\nu_1$ (for single excitons) and $\nu_2$ (for biexcitons). The transitions from two- to one-exciton states are guided by the amplitude $\Gamma^\sigma_{\nu_2\nu_1}$, which is given by a sum over all exciton-light couplings weighted by the relevant overlap between single- and two-exciton wave functions. The formalism can potentially be expanded to include the decay signatures of higher-order exciton complexes, provided that the corresponding energies and wave functions are known.
The PL signals stemming from single-/biexciton decay in Eq. \ref{eq:PL_final} can be used to estimate the life times of the respective quasi-particles, which is discussed further below. 

\section*{Results}
\begin{figure}[ht!]
\includegraphics[width=\columnwidth]{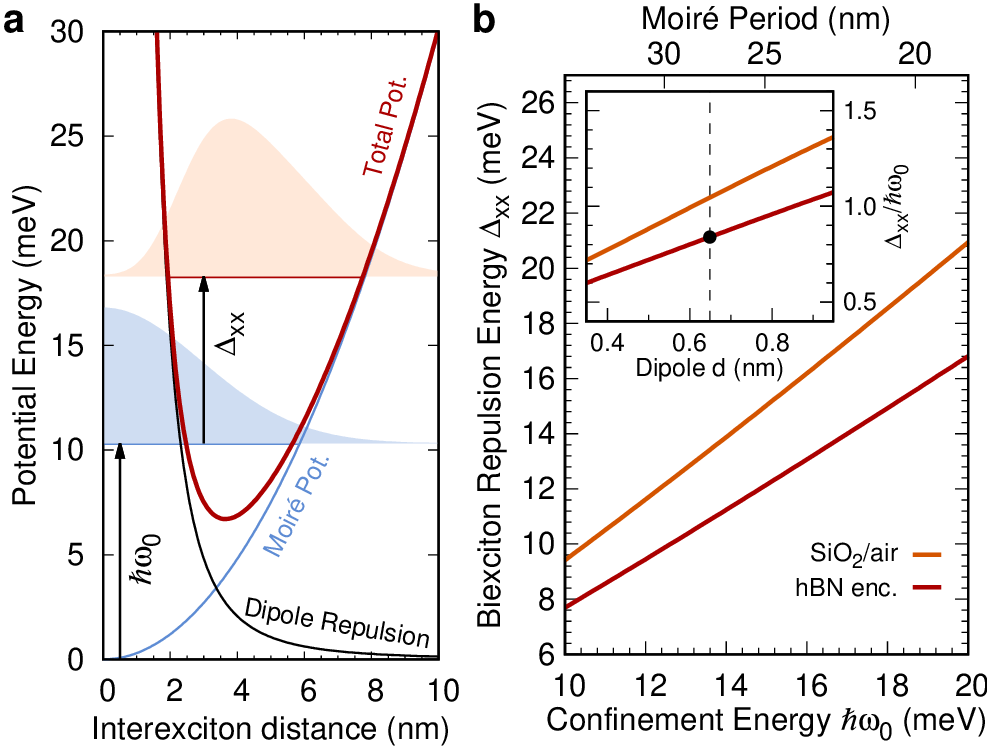}
\caption{{\bf{a}} Components of the potential acting on the relative motion between two excitons within a moir\'e trap at 0.5$^\circ$ twist angle. The moir\'e potential (blue) is forcing both particles into the potential minimum, effectively decreasing the exciton distance. The exciton-exciton potential (black line) is strongly repulsive, leading to a total potential (red) that has a minimum at a finite distance. The ground state energy (wave functions) with and without repulsion are indicated as blue and red lines (shaded curves) respectively. {\bf{b}} Biexciton repulsion energy $\Delta_\text{xx}$ as a function of confinement energy (twist angles in the range 0.5$^\circ$-1$^\circ$) on different substrates (dielectric constants). For smaller periods,  excitons are more strongly confined leading to a larger repulsion. The slope of this curve characterizes the interaction strength, which only depends on the effective dipole of the exciton (inset).}
\label{fig:prop} 
\end{figure}

Based on the microscopic model introduced above, we calculate the luminescence spectrum of moir\'e trapped excitons for different densities corresponding to filling factors smaller then three. In the first step, we analyse the energy and wave function of the repulsive biexciton with special focus on the relative motion between two trapped excitons. In Fig. \ref{fig:prop}a we illustrate the different components of the potential energy depending on the relative coordinate between two excitons, here evaluated for the specific case of hBN-encapsulated MoSe$2$/WSe$_2$ bilayers at a twist angle of $\theta_\text{twist}=0.5^\circ$. Both excitons are forced into the minimum of the moir\'e potential (blue parabola), such that the particles are effectively pushed towards each other. Without the repulsive exciton-exciton interaction, the wave function would be given by the ground state of the harmonic oscillator (blue shaded curve), while the ground state energy in 2D is given by the confinement energy $\hbar\omega_0$.

If we now additionally account for the strong dipole-dipole repulsion between the interlayer excitons (black curve), the wave function is significantly altered (red-shaded curve). The total potential (red) now has a minimum at a finite distance and diverges at zero. Consequently, the wave function exhibits a strong dip at vanishing exciton distance, which has important consequences for the interaction energy and biexciton life times. It is important to note here, that the vanishing probability at zero distance is justifying the negligence of the electron-electron/hole-hole Coulomb exchange interaction. If the biexciton wave function would contain significant contributions at interexcitonic distances on the range of the exciton Bohr radius ($\sim1.5$nm \cite{schmitt2022formation}), there would be important exchange corrections resulting from the fermionic nature of the electrons and holes \cite{shahnazaryan2017exciton}. However, the strong repulsive (direct) interaction is keeping the interlayer excitons at large enough distances to omit exchange correction terms.

We define the energetic shift between the non-interacting and the interacting case as the biexciton repulsion energy $\Delta_\text{xx}$ and show its twist-angle dependence in Fig. \ref{fig:prop}b. When neglecting the slight mismatch between monolayer lattice constants $a_0$, the moir\'e period $a_m$ is approximately inversely proportional to the twist-angle $a_m\approx a_0/\theta$. Consequently, the confinement energy scales linearly with the twist angle \cite{brem2020tunable}, cf. Eq. \ref{eq:confinement_energy}. At 1$^\circ$ the calculated confinement energy is $\sim20$meV, which agrees well with the equidistant separation between moir\'e resonances found in PL measurements \cite{tran2019evidence}. The corresponding repulsion energy for the hBN-encapsulated case $\Delta_\text{xx}\approx 17$meV is significantly smaller then the on-site energy, computed in the Hubbard model for the same system \cite{gotting2022moire}. In a Hubbard model, the interaction energies are determined in perturbation theory with respect to the non-interacting wave function (blue shaded curve in Fig. \ref{fig:prop}a), which strongly overestimates the probability to find excitons at zero distance and therefore also the repulsion energy. 
Within the limits of our model, we find a linear dependence between the twist angle (confinement energy) and repulsion energy. This agrees well with the linear dependence recently found in PL measurements in a similar material (WSe$_2$/WS$_2$) \cite{park2023dipole}, where the absolute repulsion energies found in the experiment ($\Delta_\text{xx}\sim 35$meV at $a_m\sim 8$nm) are also comparable with our predictions. The slope of this graph $\Delta_\text{xx}/\hbar\omega_0$ is characteristic for the strength of the exciton-exciton interaction and scales linearly with the effective dipole of the exciton (in this work $d\approx 0.65nm$ \cite{wu2018theory}), which is shown in the inset of Fig. \ref{fig:prop}b. Since the energy levels of single and biexcitons are proportional to the confinement energy, the  considerations in this work can be made independent of the twist angle, by referring all energies to $\hbar\omega_0$. Moreover, we focus on the most common scenario of hBN-encapsulated MoSe$2$/WSe$_2$ bilayers. 

\begin{figure}[t!]
\includegraphics[width=\columnwidth]{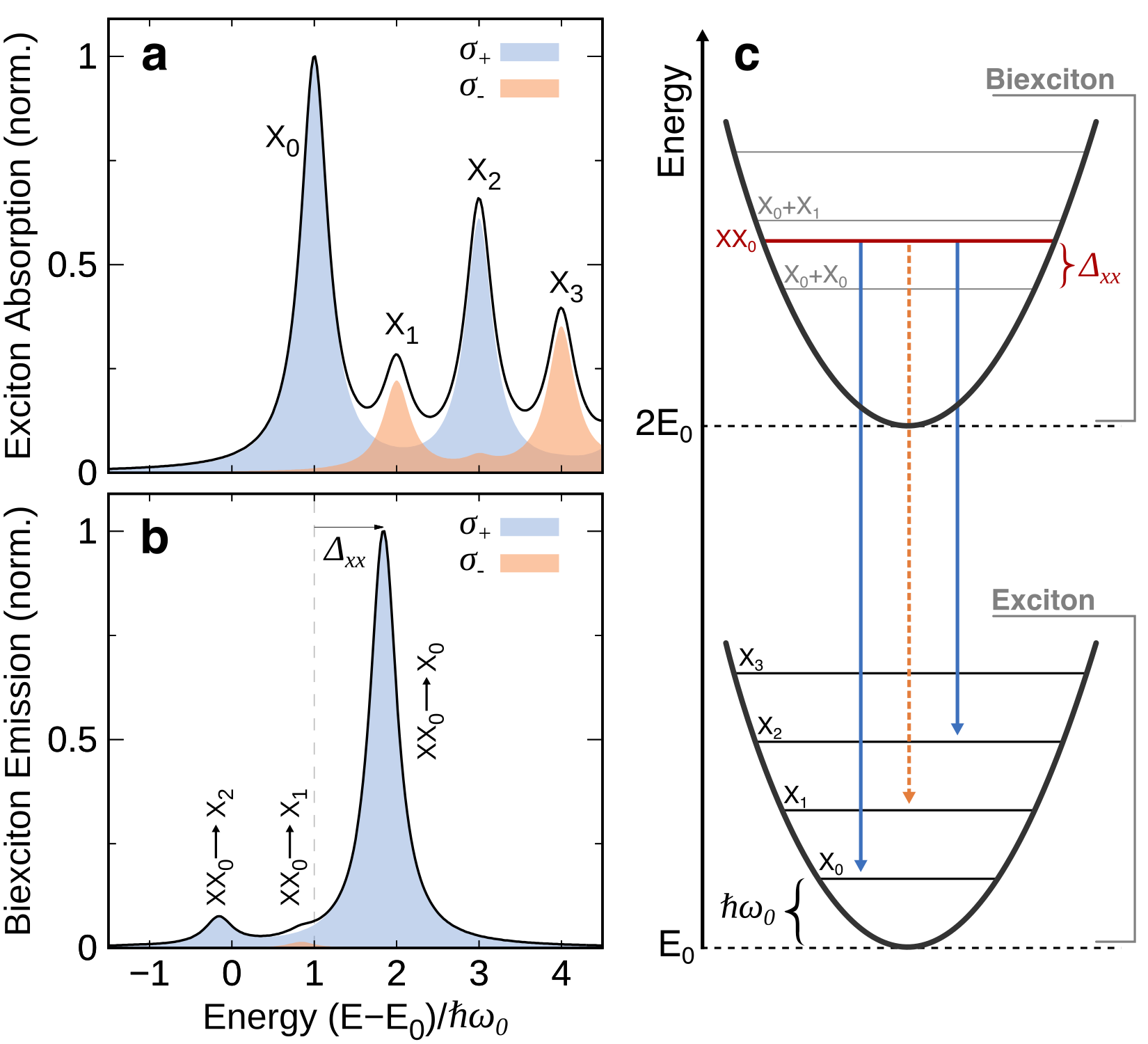}
\caption{{\bf{a}} Absorption spectrum of interlayer excitons. The ladder of moir\'e confined states X$_i$ gives rise to a series of peaks, with alternating selection rules for circularly polarized light resulting from the angular momentum of the respective state. {\bf{b}} PL spectrum of the biexciton ground state XX$_0$, exhibiting a main resonance at one repulsion energy above the single exciton ground state X$_0$. Transitions to excited single exciton states appear as side features below the main resonance. {\bf{c}} Schematic illustration of exciton and biexciton energies and possible biexciton decay channels. All energies are referred to the lowest interlayer exciton energy $E_0$, corresponding to the minimum of the moir\'e potential.}
\label{fig:res} 
\end{figure}

In the next step, we consider the radiative decay channels of the biexciton. These can be interpreted as the recombination of one of the electron-hole pairs forming the biexciton leaving behind the remainder. Therefore, we first discuss the optical activity of single moir\'e excitons, which was already analysed in previous works \cite{wu2018theory, yu2017moire, brem2020tunable}. Figure \ref{fig:res}a shows the absorption spectrum of interlayer excitons, exhibiting a series of moir\'e resonances. These peaks correspond to the series of localized states trapped in the (harmonic) potential minimum, i.e. with equidistant energy spacing $\hbar\omega_0$. Their oscillator strength and selection rules thereby result from the orbital character of the excitonic center-of-mass motion as well as from the single-particle selection rules for transitions between valence and conduction band. In TMD monolayers strict selection rules hold for excitations with circularly polarized light at the K-point (circular dichroism) \cite{cao2012valley, mak2012control}. In contrast, in moir\'e superlattices, the K-point selection rules for circularly polarized light strongly depend on the spatial position within the moir\'e pattern. In the Appendix, we provide an analytical expression for the optical selection rules for moir\'e excitons. Within our model, we find alternating circular polarizations for the series of moir\'e resonances, resulting from different angular momenta $l$ of the corresponding exciton orbitals, e.g. $l=0$ for the ground state and $l=1$ for the first excited state. This spectrum is in excellent agreement with the full evaluation of excitonic Bloch states in the moir\'e superlattice \cite{brem2020tunable} as well as polarization resolved PL measurements \cite{tran2019evidence}.     

In Fig. \ref{fig:res}b we now compare the PL signal resulting from the radiative decay of the biexciton ground state. Here, we find a main resonance at one repulsion energy above the single exciton ground state accompanied by several side features at lower energies. The microscopic origin of this spectrum can be understood by considering the energy diagram of single- and biexciton states, illustrated in Fig. \ref{fig:res}c. The single exciton levels are given by the regular ladder of equidistant harmonic oscillator states with a ground state X$_0$ at $E_0+\hbar\omega_0$. For a non-interacting biexciton (X$_0$+X$_0$), the ground state energy would thus be given by $2(E_0+\hbar\omega_0)$. In contrast, the energy of the repulsive biexciton XX$_0$ is up-shifted by $\Delta_\text{xx}$. This repulsion energy is additionally released when the biexciton decays into single exciton states, where in principle also excited orbitals (such as X$_1$) are possible final states of the remaining exciton. However, the amplitudes of the bi- to single-exciton transitions depend on the wave function overlap of initial and final states, which is the largest for XX$_0\rightarrow$ X$_0$ and much smaller e.g. for XX$_0\rightarrow$ X$_1$ due to a mismatch in angular momenta. In general, the transition amplitude towards excited single exciton states becomes larger for strong exciton-exciton interactions, since the latter lead to larger inter-exciton distance and consequently an increased overlap with large single particle orbitals.
\begin{figure*}[ht!]
\includegraphics[width=\textwidth]{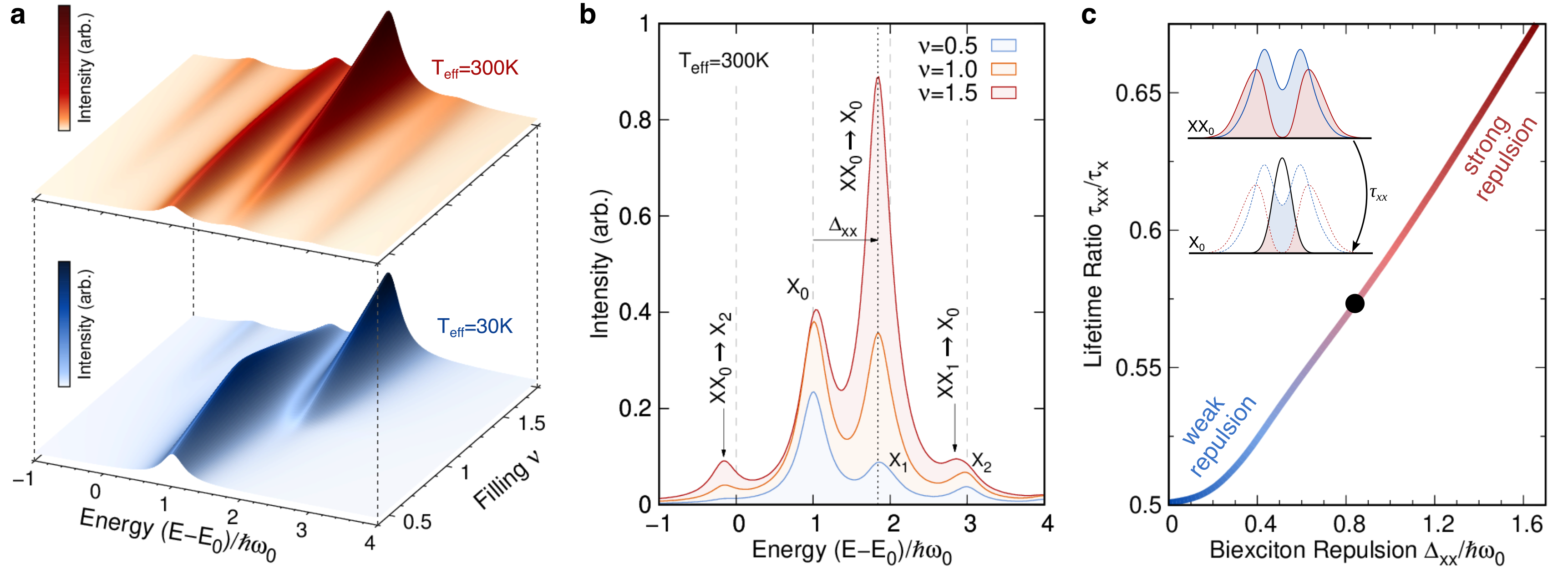}
\caption{ {\bf{a}} PL spectrum as function of filling factor for two representative effective temperatures. At fillings $>1$ the spectrum transits from a single- to a biexciton dominated emission. At larger effective temperatures, biexciton peaks appear at lower fillings and excited state features emerge. {\bf{b}} Analysis of the PL at T$_\text{eff}=300K$ and different fillings. At low filling, the lowest moir\'e resonances (X$_0$, X$_1$, X$_2$) of the moir\'e exciton ladder are visible due to a finite thermal occupation of excited single exciton states. At increased fillings, the biexciton's main decay channel XX$_0 \rightarrow$ X$_0$ becomes dominant and side peaks involving excited single-/excited biexciton states emerge at lower/higher energies (cf. Fig. \ref{fig:res}c). {\bf{c}} Ratio of the biexciton- ($\tau_\text{xx}$) and single exciton lifetime ($\tau_X$) as a function of the twist angle independent repulsion characterized by $\Delta_\text{xx}/\hbar\omega_0$. For a strong exciton-exciton repulsion, the interexciton distance becomes larger than the single exciton confinement length, leading to a decreased wave function overlap (inset) and an increased biexciton lifetime.}
\label{fig:wide} 
\end{figure*}
In PL experiments, the optically injected exciton population relaxes into a thermodynamic mixture of single- and biexciton states and the emission signal is a superposition of their respective signatures. Here it is important to note that the biexciton repulsion energy is similar to the confinement energy itself, such that the main biexciton peak appears at almost the same energy as the first excited single exciton orbital. In general, the PL spectrum at cryogenic temperatures should only exhibit signals from the lowest (single exciton) moir\'e states, since higher-order orbitals should be unoccupied ($\Delta E \sim20$meV). Nonetheless, multiple moir\'e peaks with comparable energy separations have been observed in PL experiments at cryogenic temperatures and low density \cite{tran2019evidence}. This indicates that the optically injected exciton population in experiments remains in a non-equilibrium distribution with larger mean energies as expected from the temperature of the cooled atomic lattice. Therefore, it is not trivial to decide if peaks appearing above X$_0$ are associated to excited states X$_1$ or biexcitons XX$_0$. However, a density dependent analysis of the PL spectrum can help to distinguish the microsopic origin of these peaks.

Figure \ref{fig:wide}a shows the calculated PL signal at different exciton densities, here characterized by the filling factor $n_x/n_\text{moir\'e}$, i.e. the mean number of excitons per moir\'e cell. It is important to note, that in our model we use statistical weights according to a thermal equilibrium. As discussed above, the optically injected exciton population in experiments might remain in a non-equilibrium distribution for large times. To account for this, we consider different effective exciton temperatures $T_\text{eff}$, to simulate a larger statistical spread in energies and particle numbers. Therefore, in Fig. \ref{fig:wide}a we compare the exemplary cases of high (300K, top) and low temperatures (30, bottom). In both cases we observe a density driven transfer of signal strength from the X$_0$ peak (at $E_0+\hbar\omega_0$) at low density to the XX$_0$ peak (at $E_0+\hbar\omega_0+\Delta_\text{xx}$) at fillings larger then one. In the case of a larger effective temperature, there is also a larger fluctuation in the number of excitons per cite, which e.g. can also simulate the case of spatial inhomogenities in the exciton density. In this case, biexciton features already appear at lower \textit{mean} numbers per site and on the opposite, the single exciton peak prevails at larger filling factors. Moreover, the larger energy fluctuations at higher $T_\text{eff}$ also give rise to more intense contributions of different excited states. 

In Fig. \ref{fig:wide}b we analyse the multi-peak spectrum appearing for $T_\text{eff}=300$K for three characteristic fillings $\nu$. At low filling $\nu=0.5$ we see three peaks corresponding to the lowest three single exciton orbitals X$_0$, X$_1$, X$_2$ (cf. \ref{fig:res}c), with decreasing amplitudes due to the small thermal occupation of the excited states. However, already here, there is a significant probability to form biexcitons, whose signal is superimposing the X$_1$ peak. When the density is increased to $\nu=1$, the single exciton emission features reach their maximum intensity and biexciton features start to become dominant. At filling one it is still most likely to find one exciton per cell, however, the biexciton peak has almost reached the same intensity as X$_0$. This results from the shorter biexciton lifetime, which is further discussed below. Apart from the main biexciton peak, the transition to the excited state XX$_0\rightarrow$ X$_2$ becomes clearly visible \textit{below} the single exciton ground state X$_0$. This unintuitive signature of the biexciton interestingly does not result from a bound attractive biexciton, but is in contrary a result of strong exciton-exciton repulsion as discussed above. In addition to this surprising feature of the repulsive biexciton, we also find a density enhanced emission at about $E_0+3\hbar\omega_0$. This peak stems from the first excited biexciton state XX$_1$, which can also become occupied at increased temperatures and produces its main peak via the decay XX$_1\rightarrow$ X$_0$.

Our predictions show that depending on the quantum structure of single- and biexciton states as well as the effective temperature of the system, the PL spectrum of moir\'e excitons can involve a large number of possible radiative decay channels, where single and biexciton features can be spectrally superimposed. 
The density dependent increase of the relative peak intensity of XX$_0$ compared to X$_0$ is a good indicator to distinguish the biexciton from the excited state X$_1$. However, a recent work \cite{tan2022signature} has also argued that exciton-exciton scattering at increased densities can lead to a more pronounced non-equilibrium distribution favouring the occupation of excited orbitals. To further distinguish XX$_0$ from X$_1$ features, we therefore analyse the life time of the biexciton peak based on our microscopic model. The lifetime ratio of the bi- and single exciton only depends on the interaction strength, i.e. the repulsion parameters $\Delta_\text{xx}/\hbar\omega_0$, which is illustrated in Fig. \ref{fig:wide}c. Here, the different repulsion parameters have been achieved by artificially varying the effective dipole of the interlayer exciton.

For vanishing interaction, the biexciton state decays twice as fast as the single exciton, since here the recombination probabilities of both particles forming the biexciton simply add up. When the interaction strength is increased, the life time of the biexciton becomes longer and we find a large regime in which the life time scales linearly with the repulsion strength. The microscopic origin of this behaviour is illustrated in the inset of Fig. \ref{fig:wide}c. When enhancing the exciton-exciton interaction, the probability to find excitons at zero distance is increasingly depleted. Consequently, the probability of finding a particle in the minimum of the moir\'e potential becomes quenched. With that however, the wave function overlap with the optically most active single exciton states (especially X$_0$) becomes decreased giving rise to smaller biexciton decay rates. Therefore, the lifetime ratio $\tau_\text{xx}/\tau_\text{x} > 0.5$ is a characteristic indicator for a repulsive biexciton peak and the specific value of this ratio can be used to estimate the excitonic interaction strength.

\section*{Summary}
In this work, we have gained important qualitative insights about the recombination dynamics of repulsive biexcitons trapped in moir\'e super lattices. Our analysis of the biexciton energy and the impact of the biexciton wave function on optical properties provides important tools for the interpretation of experimental results. We have highlighted several aspects that might become important in the analysis of density dependent moir\'e exciton PL spectra. First, we show that the twist angle dependence of the exciton-biexciton splitting can be used to compute the effective dipole of the interlayer exciton. Second, we show that within the experimental non-equilibrium conditions, the main biexciton feature might be superimposed by excited single particle states, but our model predicts additional unambiguous spectral features of the repulsive biexciton \textit{below} the single exciton ground state. Finally, we show that the lifetime of the biexciton is closely related to the life time of the single exciton and the life time ratio can be utilized to distinguish biexciton peaks from other signatures and, most importantly, it provides information about the strength of excitonic interaction in the system. 

We have evaluated our model for the specific case of twisted hBN-encapsulated MoSe$_2$/WSe$_2$ heterostructures, but our theoretical framework can be applied to other systems of interacting bosonic lattices, in particular involving different interaction and trapping potentials. In particular, the presence of momentum-indirect dark excitons \cite{wallauer2021momentum} or the modification of the moir\'e potential due to atomic reconstruction \cite{weston2020atomic, zhao2023excitons} could be added to the model. Moreover, our theory can be extended to include larger particle clusters for the simulation of stronger excitations conditions, and can also be adjusted to include other spin-valley configurations to model different polarizations of the excitation pulse. Finally, instead of choosing statistical weights according to a thermal equilibrium, other probability distributions can be chosen to simulate PL spectra in strong non-equilibrium conditions and/or spatial inhomogenities in energies/particle numbers e.g. due to inhomogeneous excitation and disorder \cite{rosati2021electron}. 
Overall, our work provides theoretical tools and novel insights into the optical properties of moir\'e biexcitons that will guide future theoretical and experimental studies.

\begin{acknowledgments}
We acknowledge support from Deutsche Forschungsgemeinschaft (DFG) via SFB 1083 (Project B9).
\end{acknowledgments}

\section*{Appendix}
\section{Photon Emission from a Few-Body Exciton System}
The starting point of the theoretical model is the Hamiltonian of interacting excitons subject to a periodic moir\'e potential and coupled to the quantized electromagnetic (photon-) field,
\begin{eqnarray} \label{eq:H_momentum}
H&=&H_\text{x}+H_\text{pt}+H_\text{int}\\
H_\text{x}&=&\sum_\vc{Q}\varepsilon_\vc{Q}b^\dagger_\vc{Q}b_\vc{Q}+\sum_\vc{Q,q} M_\vc{q} b^\dagger_\vc{Q+q}b_\vc{Q} \nonumber\\
&\;&\;+\sum_\vc{Q,Q',q} V_\vc{q} b^\dagger_\vc{Q+q}b^\dagger_\vc{Q'-q}b_\vc{Q'}b_\vc{Q} \\
H_\text{pt}&=&\sum_{\sigma\vc{k}}\hbar \omega_{\sigma\vc{k}} c^\dagger_{\sigma\vc{k}}c_{\sigma\vc{k}}\\
H_\text{int}&=&\sum_{\sigma\vc{k,Q}}g_{\sigma\vc{k}}(\vc{Q})  c^\dagger_{\sigma\vc{k}}b_\vc{Q} +\text{h.c.}
\end{eqnarray}
where $b^\dagger_\vc{Q}$ creates excitons with a center-of-mass momentum $\vc{Q}$ with the dispersion $\varepsilon_\vc{Q}=\hbar^2Q^2/(2m_x)$ ($m_x \approx 0.86 m_0$ \cite{kormanyos2015k}) and $c^\dagger_{\sigma\vc{k}}$ creates photons with polarization $\sigma$, momentum $\vc{k}$ and dispersion $\omega_{\sigma\vc{k}}=ck$. The exciton-exciton potential $V_\vc{q}$, the moir\'e potential $M_\vc{q}$ and the exciton-photon matrix element $g_{\sigma\vc{k}}(\vc{Q})$ are defined and discussed in the sections below. In particular, in this work we focus on interlayer excitons at the K-point of hBN encapsulated MoSe$_2$/WSe$_2$ heterostructures.
As discussed in the main text, the excitonic part of the system only contains a few particles such that its many-body eigenstates can be precisely determined, i.e.   
\begin{eqnarray} \label{eq:X-eigenstates}
H_\text{x}\ket{\nu}&=&E_{\nu}\ket{\nu}\\
\ket{\nu}&=&\sum_{\vc{Q}_1..\vc{Q}_{N_\nu}} \Psi^\nu_{\vc{Q}_1..\vc{Q}_{N_\nu}} b^\dagger_{\vc{Q}_1}..b^\dagger_{\vc{Q}_{N_\nu}}\ket{0},
\end{eqnarray}
where $\nu$ is a compound index containing the number of particles $N_\nu$ and all quantum numbers determining the configuration of the $N_\nu$-body state. The eigenstates of the photonic system are simply given by Fock states,
\begin{eqnarray} \label{eq:Pt-eigenstates}
H_\text{pt}\ket{\mu}&=&\sum_{\sigma \vc{k}} \hbar \omega_{\sigma\vc{k}} n^\mu_{\sigma\vc{k}} \ket{\mu}\\
\ket{\mu}&=&\ket{n^\mu_{\sigma\vc{k}_1},n^\mu_{\sigma\vc{k}_2}...}=\prod_{\sigma\vc{k}}\dfrac{1}{\sqrt{n^\mu_{\sigma\vc{k}}!}}( c^\dagger_{\sigma\vc{k}})^{n^\mu_{\sigma\vc{k}}}\ket{0} \nonumber
\end{eqnarray}
where the compound $\mu$ now contains the occupation numbers $n^\mu_{\sigma\vc{k}}$ of all available modes. The dynamics of the systems density matrix $\rho$ is given by the von Neumann equation, which in the basis $\ket{\alpha}=\ket{\nu_\alpha,\mu_\alpha}$ reads,
\begin{eqnarray} \label{eq:von-Neumann}
\partial_t\rho_{\alpha\beta}&=&\dfrac{i}{\hbar}(E_\beta-E_\alpha)\rho_{\alpha\beta}\nonumber\\
&\;&-\dfrac{i}{\hbar} \sum_{\gamma}(W_{\alpha\gamma}\rho_{\gamma\beta} - W_{\gamma\beta}\rho_{\gamma\beta})
\end{eqnarray}
\newline
with the interaction matrix $W_{\alpha\beta}=\bra{\alpha} H_\text{int}\ket{\beta}$.
We assume a weak coupling between excitons and photons, such that non-linear couplings are neglected and the equations of motion for coherences ($\alpha\neq \beta$) are solved in the adiabatic limit (Markov approximation),
\begin{eqnarray} \label{eq:adiabatic}
\rho_{\alpha\beta}=W_{\alpha\beta}\dfrac{\rho_{\beta\beta}-\rho_{\alpha\alpha}}{E_\beta-E_\alpha+i\Gamma},
\end{eqnarray}
where we have introduced a finite coherence lifetime $\Gamma$. In this limit the dynamics of the probabilities $P_\alpha=\rho_{\alpha\alpha}$ are given by Fermi's Golden rule ($\Gamma\rightarrow 0$)
\begin{eqnarray} \label{eq:FermisRule}
\partial_t P_\alpha=\dfrac{2\pi}{\hbar}\sum_\beta |W_{\alpha\beta}|^2(P_\beta-P_\alpha)\delta(E_\alpha-E_\beta)
\end{eqnarray}
With Eq. \ref{eq:FermisRule} we can now compute the photoluminescence (PL) spectrum as temporal change of the photon number $\<n_{\sigma\vc{k}}\>=tr(\rho c^\dagger_{\sigma\vc{k}}c_{\sigma\vc{k}})$, which neglecting absorption and stimulated emission processes yields,
\begin{eqnarray} \label{eq:PL}
\partial_t \<n_{\sigma\vc{k}}\>&=&\sum_{\alpha} n^{\mu_\alpha}_{\sigma\vc{k}} \partial_t P_\alpha\\
&=& \dfrac{2\pi}{\hbar}\sum_\vc{Q}|g_{\sigma\vc{k}}(\vc{Q})|^2 J_\vc{Q}(\omega_{\sigma\vc{k}})
\end{eqnarray}
where we have introduced the spectral intensity
\begin{eqnarray} \label{eq:Spectral}
J_\vc{Q}(\omega)=\sum_{\nu_\alpha\nu_\beta}|\bra{\nu_\alpha}b_\vc{Q}\ket{\nu_\beta}|^2P_{\nu_\beta}\delta(E_{\nu_\beta}-E_{\nu_\alpha}-\hbar\omega),
\end{eqnarray}
and the probability $P_\nu=\text{exp}[-(E_\nu-\mu_x N_\nu)/(k_BT)]/Z_x$, determined by the chemical potential $\mu_x$ for excitons and the corresponding partition function $Z_x$. 

\section{Moir\'e Potential and Single Exciton Eigenstates}

In this work, the periodic moir\'e potential determined with first principle calculations in previous works \cite{brem2020tunable}, is approximated around its minima, which allows for a numerically less expensive solution of the biexciton problem. For moir\'e periods large compared to the exciton Bohr radius, the internal structure of the exciton can be neglected and the periodic moir\'e potential can be parameterized via \cite{brem2020tunable},
\begin{eqnarray} \label{eq:MoirePeriod}
M_\vc{q}&=&\dfrac{1}{2}\sum^3_{n=0}\bigg( (A-iB)\delta_{\vc{q},\vc{g}_n} + (A+iB)\delta_{\vc{q},-\vc{g}_n} \bigg) \;\; \; \;\\
M(\vc{r})&=& \sum^3_{n=0}\bigg( A \cos(\vc{g}_n\vc{r}) + B \sin(\vc{g}_n\vc{r}) \bigg).
\end{eqnarray}
For interlayer excitons in MoSe$_2$/WSe$_2$ we found $A=0$ and $B\approx23$meV \cite{brem2020tunable}. Here the $\vc{g}_n=C_3^n\vc{g}_0$ are $120^\circ$ rotations of the fundamental reciprocal lattice vector, that can e.g. be determined by the difference between the fundamental reciprocal lattice vectors of the two monolayers $\vc{g}_0=\vc{G}_\text{W}-\vc{G}_\text{Mo}$. Shifting the real space coordinate system into one of the potential minima ($\vec{r}_0=a_m/\sqrt{3}(1,0)$) and performing a second order Taylor expansion yields,
\begin{eqnarray} \label{eq:MoireTaylor}
M(\vc{r})\approx \dfrac{1}{2} m_x \omega_0^2 r^2-U_0 \\
\omega_0=\dfrac{2\pi}{a_m}\sqrt{\dfrac{2U_0}{3m_x}}
\end{eqnarray}
with the offset $U_0=3\sqrt{3}B/2$, the exciton mass $m_x=m_c-m_v\approx0.86m_0$ \cite{kormanyos2015k} and the moir\'e period $a_m=4\pi/(\sqrt{3}|\vc{g}_0|)$ depending on the twist angle.

For the single exciton problem we define the eigenstates as,
\begin{eqnarray} \label{eq:SP-def}
H_x\ket{\nu^{(1)}} = E^{(1)}_\nu\ket{\nu^{(1)}} \;, \; \ket{\nu^{(1)}}=\sum_\vc{Q} \Phi^{(1)}_\nu(\vc{Q}) b^\dagger_\vc{Q}\ket{0}
\end{eqnarray}
which in real space yields,
\begin{eqnarray} \label{eq:SP-Schroed}
\bigg(-\dfrac{\hbar^2}{2m_x}\triangle + M(\vc{r})\bigg)\Phi^{(1)}_\nu(\vc{r})=E^{(1)}_\nu\Phi^{(1)}_\nu(\vc{r}).
\end{eqnarray}
Within the above derived parabolic approximation of the moir\'e potential, the solutions of Eq. \ref{eq:SP-Schroed} can be obtained analytically:
\begin{eqnarray} \label{eq:SP-Solution}
E^{(1)}_{nl}=\hbar\omega_0(2n+|l|+1)-U_0 \\
\Phi^{(1)}_{nl}(r,\phi)=\eta_{nl} \tilde{r} L^{|l|}_n(\tilde{r}^2)e^{-\tilde{r}^2/2+il\phi},
\end{eqnarray}
where as usual $n=0,1,2,...$ and $l=-n,-n+1,...,n$. Moreover, we introduced the
rescaling $\tilde{r}=\kappa r$, the inverse confinement length $\kappa^2=m_x\omega_0/\hbar$, the normalization $\eta_{nl}^2=\kappa^2 n!/(\pi(n+|l|)!)$ and the associated Laguerre polynomials $L^{\alpha}_n$.

\section{Dipole Repulsion and Biexciton Eigenstates}
In this work, we focus on interlayer excitons, which due to the spatial separation of electron and hole exhibit a permanent out of plane dipole moment. Therefore, the interaction between interlayer excitons is strongly repulsive, such that no Coulomb-bound biexcitons can exist. The dipolar repulsion between interlayer excitons is treated in a point-charge formalism adopted from Refs. \cite{schindler2008analysis} reading,
\begin{eqnarray} \label{eq:V_dipole}
V(r)=\dfrac{e^2}{4\pi\epsilon_r\epsilon_0}\bigg(\dfrac{2}{r}-\dfrac{2}{\sqrt{d^2+r^2}}\bigg),
\end{eqnarray}
\\
where $r$ is the inter-excitonic distance, $d\approx 0.65$nm \cite{wu2018theory} denotes the interlayer distance and $\epsilon_r$ is the averaged dielectric constant of the super-/substrate. Here we consider the case of an hBN encapsulated sample ($\epsilon_r=4.5$).

For the two exciton system the eigenvalue problem reads,
\begin{eqnarray} 
H_x\ket{\nu^{(2)}} = E^{(2)}_\nu\ket{\nu^{(2)}} \;, \label{eq:TP-def}\\
\ket{\nu^{(2)}}=\dfrac{1}{\sqrt{2}}\sum_\vc{Q,Q'} \Phi^{(2)}_\nu(\vc{Q},\vc{Q}') b^\dagger_\vc{Q}b^\dagger_\vc{Q'}\ket{0},
\end{eqnarray}
where the bosonic commutation relations enforce $\Phi^{(2)}_\nu(\vc{Q},\vc{Q}')=\Phi^{(2)}_\nu(\vc{Q}',\vc{Q})$ for the biexciton wave function. Following Eq. \ref{eq:TP-def} the latter needs to fulfill the real space Schrödinger equation,
\begin{eqnarray} \label{eq:TP-Schroed}
\bigg( &-&\dfrac{\hbar^2}{4m_x}\triangle_\vc{R} - \dfrac{\hbar^2}{m_x}\triangle_\vc{r} + V(\vc{r}) + M(\vc{R}+\vc{r}/2)\nonumber\\
 &+&M(\vc{R}-\vc{r}/2) \bigg)\hat{\Phi}^{(2)}_\nu(\vc{r},\vc{R})=E^{(2)}_\nu\hat{\Phi}^{(2)}_\nu(\vc{r},\vc{R}),
\end{eqnarray}
where we have introduced relative ($\vc{r}$) and center-of-mass coordinates ($\vc{R}$), i.e. $\hat{\Phi}^{(2)}_\nu(\vc{r},\vc{R})=\Phi^{(2)}_\nu(\vc{R}+\vc{r}/2,\vc{R}-\vc{r}/2)$. Within the parabolic approximation of the moir\'e potential (Eq. \ref{eq:MoireTaylor}) we can separated the relative and center of mass coordinates, $\hat{\Phi}^{(2)}_\nu(\vc{r},\vc{R})=\psi^{(r)}(\vc{r})\Psi^{(c)}(\vc{R})$, where the center-of-mass motion ($c$) is an unperturbed harmonic oscillation with the total mass $2m_x$, i.e.
\begin{eqnarray} 
E^{(c)}_{n_cl_c}=\hbar\omega_0(2n_c+|l_c|+1) \\
\Psi^{(c)}_{n_cl_c}(\vc{R})=\sqrt{2} \Phi^{(1)}_{n_cl_c}(\sqrt{2}\vc{R}),\label{eq:CoM-Solution}
\end{eqnarray}
where again $n_c=0,1,2,...$ and $l_c=-n_c,-n_c+1,...,n_c$. The factor $\sqrt{2}$ in Eq. \ref{eq:CoM-Solution} accounts for the change in the confinement length $1/\kappa$ due to the increased mass. For the relative motion, the total potential is given by the moiré trap and the repulsive exciton-exciton interaction $V^{(r)}(\vc{r})=V(\vc{r})+1/4m_x\omega_0^2r^2$. We solve the corresponding eigenvalue problem by expanding the relative wave function in terms of the unperturbed harmonic oscillators (i.e. without repulsion) with quantum numbers $(n_r,l_r)$. Here the radial symmetry of the interaction potential forbids the mixing of different angular momentum states, such that $l_r$ is still a good quantum number. Hence we find, 
\begin{eqnarray} 
\psi^{(r)}_{\nu_rl_r}(\vc{r})=\dfrac{1}{\sqrt{2}}\sum^\infty_{n_r=|l_r|} c_{\nu_rl_r}(n_r) \Phi^{(1)}_{n_rl_r}(\vc{r}/\sqrt{2}),
\end{eqnarray}
where $l_r$ needs to be an even number since the bosonic commutation relations enforce $\psi^{(r)}(\vc{r})=\psi^{(r)}(-\vc{r})$ \cite{mujal2017quantum}. The mixing coefficients have to fulfill the equation,
\begin{eqnarray} 
\sum_{n_r'}\bigg(\hbar\omega_0(2n_r'+|l_r|+1)\delta_{n_rn_r'}+\tilde{V}^{l_r}_{n_rn_r'}\bigg)c_{\nu_rl_r}(n_r')\nonumber\\
=E^{(r)}_{\nu_rl_r} c_{\nu_rl_r}(n_r) \\
\tilde{V}^{l_r}_{n_rn_r'}=\dfrac{1}{2}\int d^2 r \Phi^{(1)\ast}_{n_rl_r}(\vc{r}/\sqrt{2}) V(r)  \Phi^{(1)}_{n_r'l_r}(\vc{r}/\sqrt{2})
\end{eqnarray}
which finally yields the total energy of the biexciton $E^{(2)}_{n_cl_c\nu_rl_r}=E^{(c)}_{n_cl_c}+E^{(r)}_{\nu_rl_r}-2U_0$.

\section{Evaluation of Emission Spectra and Lifetimes}
In the following we evaluate the emission spectrum Eq. \ref{eq:PL} by considering the main contributions to the spectral function Eq. \ref{eq:Spectral} at low filling factors. In the first step we analyse the decay spectrum of single excitons states, i.e. transitions $\ket{\nu^{(1)}}\rightarrow \ket{0}$. In a previous work \cite{brem2020tunable}, we have derived the exciton-photon matrix element for interlayer excitons at the K-point in a twisted TMD bilayer based on a tight-binding model, yielding
\begin{eqnarray}
    g_{\sigma \vc{k}}(\vc{Q})= g_0 \vc{e_{\sigma \vc{k}}} \sum_{n=0}^2 C_3^n\dfrac{\vc{K}_v}{|\vc{K}_v|}e^{i 2\pi/3 n} \delta_{\vc{Q},C_3^n(\vc{K}_c-\vc{K}_v)},
\end{eqnarray}
where $g_0$ is a constant, which determines the overall oscillator strength of the interlayer exciton, $\vc{e_{\sigma \vc{k}}}$ is the polarization vector of the photon and $\vc{K}_{c(v)}$ is the K-point of the layer hosting the conduction (valence) band. We simplify Eq. \ref{eq:PL} by transforming it into the single exciton eigenbasis,
\begin{eqnarray}
 b^\dagger_{\nu}=\sum_\vc{Q} \Phi^{(1)}_\nu(\vc{Q}) b^\dagger_\vc{Q},
\end{eqnarray}
yielding
\begin{eqnarray} \label{eq:PL_eigen}
\partial_t \<n_{\sigma\vc{k}}\>=\dfrac{2\pi}{\hbar}\sum_\nu|g_{\sigma\vc{k}}(\nu)|^2 J_\nu(\omega_{\sigma\vc{k}})
\end{eqnarray}
and the new definitions
\begin{eqnarray} \label{eq:Spectral_eigen}
g_{\sigma\vc{k}}(\nu)&=&\sum_\vc{Q}   g_{\sigma \vc{k}}(\vc{Q}) \Phi^{(1)}_\nu(\vc{Q}) \\
J_\nu(\omega)&=&\sum_{\nu_\alpha\nu_\beta}|\bra{\nu_\alpha}b_\nu\ket{\nu_\beta}|^2P_{\nu_\beta}\delta(E_{\nu_\beta}-E_{\nu_\alpha}-\hbar\omega), \nonumber
\end{eqnarray}
For simplicity, we focus on the PL emitted in perpendicular direction with respect to the monolayer, such that the two possible polarizations are given by Jones vectors in the monolayer plane $\vc{e}_{\sigma}=1/\sqrt{2}(1,i\sigma)$ with $\sigma=\pm 1$ for circularly left/right polarized light. Moreover, we can split of the angle dependence the exciton wave function $\Phi^{(1)}_\nu(\vc{Q})=\tilde{\Phi}^{(1)}_\nu(|\vc{Q}|)\exp(il_\nu\phi_\vc{Q})$, which yields an analytical formula for the optical selection rules in perpendicular geometry,
\begin{eqnarray} \label{eq:selection_rules}
g^\perp_{\sigma}(\nu)&=&g_0 \tilde{\Phi}^{(1)}_\nu(|\vc{K}_c-\vc{K}_v|)f(\sigma+l_\nu-1) \\
f(n)&=&1+2\cos(\dfrac{2\pi}{3}n)=\begin{cases}
    3 & \text{for } n=3m  ; \; m \in \mathbb{Z}\\
     0 & \text{else},
   \end{cases} \nonumber
\end{eqnarray}
We can directly read off important optical selection rules for moir\'e trapped excitons: The lowest moir\'e exciton ($l_\nu=0$) only couples to circular right polarized light ($\sigma=+1$). Note that the above analysis has been performed for excitons at the K-point, whereas excitons in the K'-valley will have opposite selection rules reflecting the well known circular dichroism in TMDs \cite{cao2012valley, mak2012control}. In contrast, the first excited states in 2D are p-type ($l_\nu=\pm 1$), such that one of them is dark and the other one couples to  circular left polarized light ($\sigma=-1$).  

Next, we decompose the spectral function into contributions stemming from the decay of exciton complexes with different particle numbers, where we first focus on the single exciton contribution, i.e.
\begin{eqnarray} \label{eq:Spectral_single}
J^{(1)}_\nu(\omega)=\sum_{\nu'}|\bra{0}b_\nu\ket{\nu'^{(1)}}|^2P^{(1)}_{\nu'}\delta(E^{(1)}_{\nu'}-\hbar\omega),
\end{eqnarray}
Hence, for the perpendicular PL emission from single exciton states we obtain
\begin{eqnarray} \label{eq:PL_single}
I^{(1)}_\sigma(\omega)=\dfrac{2\pi}{\hbar}\sum_{\nu}|g^\perp_{\sigma}(\nu)|^2 P^{(1)}_{\nu}\delta(E^{(1)}_{\nu}-\hbar\omega),
\end{eqnarray}
corresponding to a peak at every single moir\'e exciton energy $E^{(1)}_{\nu}$, whose intensity depends on the oscillator strength $|g^\perp_{\sigma}(\nu)|^2$ and the occupation probability $P^{(1)}_{\nu}=\text{exp}[-(E^{(1)}_\nu-\mu_x)/(k_BT)]/Z_x$.
Finally, we also consider the decay spectrum of biexcitons, i.e. transitions from two-exciton to single-exciton states,
\begin{eqnarray} \label{eq:Spectral_two}
J^{(2)}_\nu(\omega)=\sum_{\nu_\alpha\nu_\beta}|\bra{\nu_\alpha^{(1)}}b_\nu\ket{\nu_\beta^{(2)}}|^2P^{(2)}_{\nu_\beta}\delta(E^{(2)}_{\nu_\beta}-E^{(1)}_{\nu_\alpha}-\hbar\omega).\nonumber
\end{eqnarray}
Here we introduce the important wave function overlap,
\begin{eqnarray} \label{eq:1-2-overlap}
\gamma^{\nu_1\nu_2}_{\nu}&=& \dfrac{1}{\sqrt{2}} \bra{\nu_1^{(1)}} b_{\nu_2}\ket{\nu^{(2)}}, \\
&=& \sum_{\vc{Q,Q'}} \Phi^{(1)\ast}_{\nu_1}(\vc{Q})\Phi^{(1)\ast}_{\nu_2}(\vc{Q})\Phi^{(2)}_{\nu}(\vc{Q,Q'})
\end{eqnarray}
such that the perpendicular PL reads,
\begin{eqnarray} \label{eq:PL_two}
I^{(2)}_\sigma(\omega)&=&\dfrac{2\pi}{\hbar}\sum_{\nu\nu_1}\Gamma^\sigma_{\nu\nu_1} P^{(2)}_{\nu}\delta(E^{(2)}_{\nu}-E^{(1)}_{\nu_1}-\hbar\omega), \\
\Gamma^\sigma_{\nu\nu_1}&=&2\sum_{\nu_2} |g^\perp_{\sigma}(\nu_2)\gamma^{\nu_1\nu_2}_{\nu}|^2,
\end{eqnarray}
and the biexciton occupation probability $P^{(2)}_{\nu}=\text{exp}[-(E^{(2)}_\nu-2\mu_x)/(k_BT)]/Z_x$. Combining Eq. \ref{eq:PL_single} and \ref{eq:PL_two} the total PL spectrum for filling factors smaller then three is approximated via $I\approx I^{(1)}+I^{(2)}$. The chemical potential $\mu_x$ is then determined through the filling factor $f$ by numerically inverting the condition $f=\sum_{\nu}P^{(1)}_{\nu}+2\sum_{\nu}P^{(2)}_{\nu}$, where the two sums refer to all possible single-/two-exciton eigenstates, respectively. Finally, we can obtain access to the single and biexciton radiative lifetimes by summing over all possible decay channels, i.e. $1/\tau^{(1)}_\nu\propto \sum_\sigma |g^\perp_{\sigma}(\nu)|^2$ and $1/\tau^{(2)}_\nu\propto \sum_{\sigma\nu_1} \Gamma^\sigma_{\nu\nu_1}$. Consequently, considering e.g. the lifetime ratio $\zeta$ of the single- and biexiton ground state,
\begin{eqnarray} \label{eq:ratio}
\zeta=\tau^{(2)}_0/\tau^{(1)}_0= \dfrac{ \sum_\sigma |g^\perp_{\sigma}(0)|^2 }{ \sum_{\sigma\nu_1} \Gamma^\sigma_{0\nu_1}},
\end{eqnarray}
we obtain a quantity that is independent of the absolute oscillator strength $g_0$, but only reflects the strength of the excitonic interaction, i.e. $\zeta=0.5$ for no interaction and $\zeta \rightarrow \infty$ for increasing interaction strength, as further discussed in the main text.

\section*{References}
\bibliographystyle{achemso}
\providecommand{\latin}[1]{#1}
\makeatletter
\providecommand{\doi}
  {\begingroup\let\do\@makeother\dospecials
  \catcode`\{=1 \catcode`\}=2 \doi@aux}
\providecommand{\doi@aux}[1]{\endgroup\texttt{#1}}
\makeatother
\providecommand*\mcitethebibliography{\thebibliography}
\csname @ifundefined\endcsname{endmcitethebibliography}  {\let\endmcitethebibliography\endthebibliography}{}

\end{document}